\documentclass{elsart}

\begin{document}

\begin{frontmatter}

\title{Understanding High-Density Matter Through Analysis of Surface Spectral 
Lines and Burst Oscillations from Accreting Neutron Stars\thanksref{ack}}
\thanks[ack]{This work was supported in part by NSF grant AST~0098436 at
Maryland, by NSF grant AST~0098399 and NASA grants NAG~5-12030 and
NAG~5-8740 at Illinois, and by the funds of the Fortner Endowed Chair at
the University of Illinois.}

\author[label1,label2]{Sudip Bhattacharyya\corauthref{cor1}},
\author[label1]{M. Coleman Miller},
\author[label2]{Tod E. Strohmayer},
\author[label3]{Frederick K. Lamb},
\author[label1,label2]{Craig B. Markwardt}
\address[label1]{University of Maryland at College Park, USA}
\address[label2]{Goddard Space Flight Center, NASA, USA}
\address[label3]{University of Illinois at Urbana-Champaign, USA}
\corauth[cor1]{Code 662, NASA/GSFC, Greenbelt, MD 20771, USA; 
sudip@milkyway.gsfc.nasa.gov}

\begin{abstract}
We discuss millisecond period brightness oscillations and surface
atomic spectral lines observed during type I X-ray bursts from a
neutron star in a low mass X-ray binary system.  We show that
modeling of these phenomena can constrain models of the dense cold
matter at the cores of neutron stars. We demonstrate that, even for a
broad and asymmetric spectral line, the stellar radius-to-mass ratio
can be inferred to better than 5\%. We also fit our theoretical models
to the burst oscillation data of the low mass X-ray binary XTE
J1814-338, and find that the 90\% confidence lower limit of the
neutron star's dimensionless radius-to-mass ratio is 4.2.
\end{abstract}


\end{frontmatter}

\section{Introduction}
\label{sec1}

Understanding the properties of very high density $(\sim 10^{15}$ gm 
cm$^{-3}$, i.e., beyond nuclear density) cold matter at the cores of 
neutron stars is a fundamental problem of physics. 
Constraints on the proposed theoretical equation of state (EOS) models
of such matter
based on terrestrial experiments are difficult, because no
experiments at  such extreme densities at low temperature seem
possible. The only way to address this problem is to measure
the mass, the radius and the spin period of the same neutron star,
as for a given EOS model and for a known stellar spin period, there 
exists a unique mass vs.  radius relation for neutron stars.
Any periodic variation in the observed lightcurve will provide us with
the stellar spin period, if we can show that this periodic variation is
due to stellar spin. But mass measurements usually require fortuitous
observations of binaries, and radius estimates have historically been
plagued with systematic uncertainties (see van Kerkwijk 2004 for a
recent summary of methods). Moreover, none of these methods can
measure all three parameters of the same neutron star
that are needed to constrain EOS models effectively.

We explore instead constraints based on the study of type I X-ray bursts
from an accreting neutron star in an LMXB system. These bursts are
convincingly explained as thermonuclear flashes on the stellar surface
(Strohmayer \& Bildsten 2003, and references therein), 
and hence can give us information about the stellar
parameters.  In addition, the comparatively low magnetic field of a
neutron star in an LMXB  does not complicate the stellar emission of
photons much (and hence  keeps the modeling simple), which may not be
the case for isolated neutron stars or neutron stars in other systems.
The millisecond period brightness oscillations during type I X-ray
bursts provide us with the stellar spin period, as this phenomenon is
caused by the combination of stellar spin and an asymmetric brightness
pattern on the stellar surface (Chakrabarty et al. 2003; Strohmayer et
al. 2003). During these X-ray bursts, atomic spectral lines may be
observed from the stellar surface (as might be the case for the LMXB
EXO~0748$-$676; see Cottam, Paerels \& M\'endez 2002). When properly
identified, these lines provide the surface gravitational redshift
value, and hence the stellar radius-to-mass ratio. The remaining
stellar parameter can be obtained by detailed modeling of the
structures of the burst oscillation lightcurves, 
and broadened \& skewed (due to rapid stellar spin) 
surface atomic spectral lines. Here we calculate such theoretical models,
and fit the burst oscillation models to the data of 
the LMXB XTE J1814-338 to constrain some stellar parameters. 

\section{Model Computation}
\label{sec2}

For the computation of burst oscillation lightcurves, we assume that the
X-ray emitting region is a uniform circular hot spot on the stellar
surface. In contrast, for our calculations of surface atomic spectral
lines, we assume that the X-ray emitting portion is a belt that is
symmetric around the stellar spin axis. This is because, for a
typical spin frequency $> 10$~Hz, any hot spot on the stellar surface
will be effectively smeared into an axisymmetric belt during a typical 
integration time for spectral calculations. In both
calculations, we consider the following physical effects (first four of
these were considered by \"Ozel \& Psaltis 2003): (1) Doppler
shift due to stellar rotation, (2) special relativistic beaming, (3)
gravitational redshift, (4) light-bending, and (5) frame-dragging. To
include the effects of light-bending, we trace back the paths of the
photons (numerically, in the Kerr spacetime) from the observer to the
source using the method  described in Bhattacharyya, Bhattacharya \&
Thampan (2001).

For a given EOS model, we have the following source
parameters: two stellar structure parameters (radius-to-mass ratio and
spin frequency), one binary parameter (observer's inclination angle), two
emission region parameters (polar angle position and angular width of the
belt or the hot spot), and a parameter $n$ describing the emitted specific
intensity distribution (in the corotating frame) of the form
$I(\alpha)\propto \cos^n(\alpha)$, where $\alpha$ is the emission angle
measured from surface normal. Other stellar structure parameters (mass and
angular momentum) are found by computing the structure of the
spinning neutron star, using the formalism given by Cook, Shapiro \&
Teukolsky (1994; see also Bhattacharyya et al. 2000, Bhattacharyya, Misra
\& Thampan 2001).

\section{Results and Discussions}
\label{sec3}

We have two main results:
(1) if the stellar radius-to-mass ratio is inferred from the
line centroid (which is the geometric mean of the low-energy and high-energy
edges of the line profile), 
the corresponding error is in general less than 5\% (Bhattacharyya,
Miller \& Lamb 2004), even when the observed surface line is broad and 
asymmetric. This is the accuracy needed for
strong constraints on neutron star EOS models (Lattimer \&
Prakash 2001). Other methods to infer this stellar parameter using surface lines
(for example, using the peak energy of a line; see
\"Ozel \& Psaltis 2003) give much larger errors.
(2) The 90\% confidence lower limit of the dimensionless 
stellar radius-to-mass ratio of the LMXB XTE J1814-338 is 4.2 (see
Bhattacharyya et al. 2005).

These results show that both surface spectral lines and burst oscillations can
provide important information about the stellar parameters. These two 
phenomena were observed from the neutron star in the LMXB
EXO~0748$-$676, and gave the values of stellar spin period (Villarreal
\& Strohmayer 2004) and
radius-to-mass ratio (Cottam, Paerels \& Mendez 2002). 
However, to measure the remaining 
stellar parameter (needed to constrain EOS models),
we need larger instruments, partly due to
the low duty cycle of bursts.
But this at least indicates that these two surface
spectral and timing features can originate from the same neutron
star, and may help us constrain neutron star EOS models
effectively, when observed with large area detectors of
future generation X-ray missions, such as Constellation-X and XEUS.

\end{document}